\documentclass[showpacs,aps,pre,twocolumn]{revtex4}
\usepackage[dvips]{graphics}
\usepackage{graphicx}
\usepackage{amsfonts}
\usepackage{amssymb}
\usepackage{amsmath}
\usepackage{epsfig}

\begin{document}

\noindent
{\bf Comment on ``Efimov States and their Fano Resonances
in a Neutron-Rich Nucleus''}
\vspace{0.5cm}

In a three-body model for $^{20}$C (neutron $-$ neutron $-^{18}$C), 
Mazumdar et al.~\cite{Mazumdar} have recently
claimed that a bound Efimov state~\cite{Efimov} 
turns into a continuum resonance when the $^{19}$C binding is
varied. They also point out this result as due to the mass
difference of the particles. Such claims need to be more deeply investigated, 
as they are in apparent confront with previous studies of Efimov physics with 
three equal-mass particles~\cite{adhprc82,amadonoble}.

The Efimov effect is strictly valid in the zero-range limit~\cite{Efimov}, 
when the scattering lengths are much larger than the effective range.  
So, it is appropriate to consider zero-range models as in \cite{amorim,revs}. 
Within our renormalized zero-range approach, we show that the results obtained for the 
$n-n-^{18}$C trajectory of Efimov states are consistent with previous studies 
that have used separable interactions~\cite{adhprc82,amadonoble}. 
As the unitarity cuts in the complex energy plane come from the $n-^{19}$C 
elastic scattering and the three-body breakup, we extended the bound-state equations 
to the second energy sheet through the $n-^{19}$C elastic scattering cut,  
by using the method given in~\cite{Gloeckle}.
Our results, for varying $E_{^{19}{\rm C}}$ binding,
with the $n-n$ virtual-state energy ($=-$143 keV) and
three-body ground-state ($E^{gs}_{^{20}{\rm C}}=-$3.5 MeV) 
fixed, are shown in Table I. 
In contrast with \cite{Mazumdar}, no three-body resonances
were found.
\begin{table}[thb!]
\caption{ $|E_{^{20}{\rm C}}-E_{^{19}{\rm C}}|$
excited states for $E^{gs}_{^{20}{\rm C}}=-$3.5 MeV and varying binding
$E_{^{19}{\rm C}}$.
Indices $V$ refer to virtual states.}
\begin{tabular}{cccccccc}
\hline
$|E_{^{19}{\rm C}}| $ (keV) $=$& 200 & 190 &180 & 170 & 160 & 150 & 140 \\
$|E_{^{20}{\rm C}}-E_{^{19}{\rm C}}|$(eV) $\simeq$
& 668$^V$& 339$^V$& 122$^V$& 14$^V$& 12& 115 & 317 \\
\hline
\end{tabular}
\end{table}
We also checked that 
the $n-^{19}$C elastic scattering cross sections for
$E_{^{19}{\rm C}}=-$150, $-$180, and $-$500 keV, 
decrease monotonically~\cite{arxiv}. The $n-^{19}$C scattering
lengths  for these three cases are, respectively, 448.5, $-$411.3 and
$-$8.6 fm (Minus signs are for virtual states).
In principle, one can vary the ground-state three-body energy 
[$E^{gs}_{^{20}{\rm C}}$]; however, 
as shown in Ref.~\cite{amorim} (see also pps.
327-329 of Braaten and Hammer~\cite{revs}) 
only the ratios between the two-body energies and three-body 
ground state energy are enough to determine the first excited 
Efimov state.

By considering $n-n-c$ three-body systems with asymmetric two-body
energies, in \cite{amorim} it was mapped a parametric region defined
by the $s-$wave two-body (bound or virtual) energies, where Efimov
bound states can exist. At the critical boundary, as we increase the
two-body (bound or virtual) energies the Efimov excited energies 
enter in the second sheet of the complex energy plane as virtual or 
resonant states (see Fig. 1 of \cite{fb18}). 
When at least one of  the subsystem is bound, such as $n-^{18}$C in 
the $^{20}$C, they become virtual states, entering the second energy 
sheet through the $n-^{19}$C elastic cut.
This is in agreement with Refs.~\cite{amadonoble,adhprc82}.
In order to move the three-body energy pole from a bound directly to 
a resonant state we need a Borromean system (all subsystems unbound)~\cite{resonance}. 
Such case (see Fig.1c of \cite{fb18}) corresponds to the Innsbruck 
experiment with Cesium atoms~\cite{grimm}.

By introducing a mass asymmetry in a non-Borromean three-body system, 
without changing the energy relations, the virtual state pole cannot 
move from the negative real axis of the complex energy plane (with nonzero width) 
and become a resonance, because the analytical structure of the unitarity 
cuts remains the same.

\vspace{0.2cm}
\noindent { M.T. Yamashita$^1$, T. Frederico$^2$ and Lauro Tomio$^3$} $^{(*)}$
\vskip 0.1cm
\indent
$^1$ Campus Experimental de Itapeva, Universidade Estadual Paulista,
18409-010, Itapeva, SP, Brazil. \\
\indent
$^2$ Departamento de F\'\i sica, Instituto Tecnol\'ogico de
Aeron\'autica, 12228-900, S\~ao Jos\'e dos Campos, Brazil.\\
\indent
$^3$ Instituto de F\'\i sica   Te\'orica, Universidade
Estadual Paulista, 01405-900, S\~ao Paulo, Brazil.

{$^{(*)}$ We thank the Brazilian agencies FAPESP and CNPq for partial support.}
\end{document}